\documentclass[twocolumn]{aastex631}

\usepackage{graphicx}	
\usepackage{amsmath}	
\usepackage{amssymb}	
\usepackage{xcolor}
\usepackage{savesym}
\savesymbol{tablenum}
\usepackage{siunitx}
\restoresymbol{SIX}{tablenum}
\usepackage{array}
\usepackage{makecell}
\usepackage{cprotect}

\begin{document}

\title{Evidence for GeV Gamma-Ray Emission from Intense GRB~240529A During the Afterglow's Shallow Decay Phase}

\correspondingauthor{Kenta Terauchi}
\email{terauchi.kenta.74s@st.kyoto-u.ac.jp}

\author[0009-0002-3519-2535]{Kenta Terauchi}
\affiliation{Department of Physics, Kyoto University, Kitashirakawa Oiwake-cho, Sakyo, Kyoto, 606-8502, Japan}

\author[0000-0002-9924-9978]{Tomohiko Oka}
\affiliation{Research Organization of Science and Technology, Ritsumeikan University, 1-1-1 NojiHigashi, Kusatsu, 525-8577, Shiga, Japan}

\author[0000-0001-9064-160X]{Katsuaki Asano}
\affiliation{Institute for Cosmic Ray Research, The University of Tokyo, 5-1-5 Kashiwanoha, Kashiwa, Chiba 277-8582, Japan}



\begin{abstract}
X-ray light curves of gamma-ray burst (GRB) afterglows exhibit various features, with the shallow decay phase being particularly puzzling. 
While some studies report absence of the X-ray shallow decay for hyper-energetic GRBs, recently discovered GRB~240529A shows a clear shallow decay phase with an isotropic gamma-ray energy of \SI{2.2e54}{erg}, making it a highly unusual case compared to typical GRBs.
In order to investigate the physical mechanism of the shallow decay, we perform the \textit{Fermi}-LAT analysis of GRB~240529A along with \textit{Swift}-XRT analysis.
We find no jet break feature in the X-ray light curve and then give the lower bound of the collimation-corrected jet energy of $>10^{52}$~erg, which is close to the maximum rotational energy of a magnetar.
Our LAT data analysis reveals evidence of GeV emission with a statistical significance of $4.5\sigma$ during the shallow decay phase, which can be interpreted as the first case for hyper-energetic GRBs with a typical shallow decay phase.
The GeV to keV flux ratio is calculated to be $4.2\pm2.3$. Together with X-ray spectral index, this indicates an inverse Compton origin of the GeV emission.
Multiwavelength modeling based on time-dependent simulations tested two promising models, the energy injection and wind models.
Both models can explain the X-ray and gamma-ray data, while our modeling demonstrates that gamma-ray observations, along with future GeV--TeV observations by CTAO, will distinguish between them.


\end{abstract}



\section{Introduction} 
\label{sec:intro}

Gamma-ray bursts (GRBs) are the brightest explosions in the Universe and exhibit two types of emission phases: the prompt emission and the afterglow emission. 
While the prompt emission typically lasts few milliseconds to several thousands of seconds, displaying a complex light curve with sub-second variability, the subsequent afterglow emission can be detected for several days or longer, with its light curve exhibiting different features mostly characterized by power-law decay \citep{Kumar_2015}. 

X-ray afterglows in a large fraction of GRBs (approximately 90\%; see \citet{Yamazaki_2020, Liang_2009})show a shallow decay behavior ($f_\nu \propto T^{-0.5}$ or shallower) during initial $\sim$\SI{e3}{\second} \citep{Nousek_2006, Liang_2007}.
Although the origin of the shallow decay is still largely unknown, it is generally believed to be related to the central engine of GRBs.
Several models are proposed to explain the shallow decay such as the energy injection model \citep{Zhang_2001, Zhang_2006, Granot_Kumar_2006}, 
evolving microphysical parameters \citep{Fan_Piran_2006, Panaitescu_2006, Ioka_2006}, and the wind model \citep{Shen_2012, Dereli-Begue_2022}.
While a long activity of the central engine, such as a magnetar, has been considered for the energy injection model \citep{Zhang_2006}, \citet{2024arXiv240810750K} show that the finite thickness of the ejecta can naturally cause the evolution of the forward shock equivalent to the energy injection model. 
The wind model is also attractive. Unlike the energy injection model, it does not require a long activity for the central engine with a non-trivial evolution of the luminosity or bulk Lorentz factor \citep{Shen_2012, Dereli-Begue_2022}.


GeV–TeV gamma-ray observations can provide clues to this enigmatic shallow decay phase. 
Recently \cite{2024ApJ...970..141A} showed that GeV--TeV observations can help to determine the different models of shallow decay phase.
\cite{Dereli-Begue_2022} showed that non-detection of GeV photon can be used to test the wind model if synchrotron emission is considered, while \citet{2024ApJ...970..141A} suggests that inverse Compton emission can produce a bright GeV--TeV emission.
GeV afterglows in initial $10^3$--$10^4$\,s are detected by the Large Area Telescope (LAT; \cite{Atwood_2009}) on the \textit{Fermi} space telescope for a fraction of GRBs \citep{Ajello_2019}. 
However, \cite{Yamazaki_2020} argued that GeV--TeV GRBs have less noticeable shallow decay phase in the early X-ray afterglow: Even if the GeV--TeV GRBs have shallow decay phase, their decay indices tend to be steeper than the typical value.
Moreover, based on the results from \citet{Ding_2022}, only 3\% of their sample of the shallow decay GRBs have $E_{\rm \gamma,iso}$ larger than  \SI{e54}{erg}. This value is much lower than the fraction of GRBs with just $E_{\rm \gamma,iso} > \SI{e54}{erg}$, which is $\sim24\%$ \citep{Atteia_2017}.
\cite{Sharma_2021} argued that hyper-energetic GRBs with beam-corrected prompt energy greater than $\sim \SI{e52}{erg}$ show no shallow decay feature in their X-ray light curve.
The authors also claimed that the central engine of those hyper-energetic GRBs is a black hole rather than a magnetar, since the total burst energy would exceed the maximum possible rotational energy of the magnetar that can be converted into the GRB jet.
These results might suggest that a black hole is not related to the X-ray shallow decay. However, some studies argue that the X-ray shallow decay is mostly attributed to a black hole under an assumption of isotropic energy release from a magnetar during the shallow decay (e.g. \citet{Li_2018}).



The detection of GRB~240529A was initially reported by the Burst Alert Telescope (BAT; \citet{Barthelmy_2005}) on the \textit{Neil Gehrels Swift Observatory} (\textit{Swift}) spacecraft on 29 May 2024, 02:58:31 UT (hereafter $T_0$; \cite{GCN36556}).
The prompt duration $T_{90}$ (the time corresponding to 5\%--95\% of a burst fluence)
is $160.67 \pm 14.52$~s in the 15--\SI{350}{\keV} band. 
The time-averaged BAT spectrum is well modeled by a single power-law model with photon index $\Gamma=-1.68 \pm 0.04$, suggesting a high peak energy above 150 keV. 
The fluence in the 15--$\SI{150}{\keV}$ band is $S_{\gamma,\text{15--150}}=\SI{2.1(0.0)e-5}{erg\,\cm^{-2}}$ \citep{GCN36566}.
The prompt emission was also detected by the \textit{AstroSat} satellite \citep{GCN36560}, the Hard X-ray Modulation Telescope \citep{GCN36578}, \textit{Mars Odyssey} \citep{GCN36583}, and Konus-\textit{Wind} \citep{GCN36584}.
There was no onboard trigger of the prompt emission by the \textit{Fermi} Gamma-Ray Burst Monitor.

\begin{figure*}[ht!]
     \epsscale{1.1}
    \plottwo{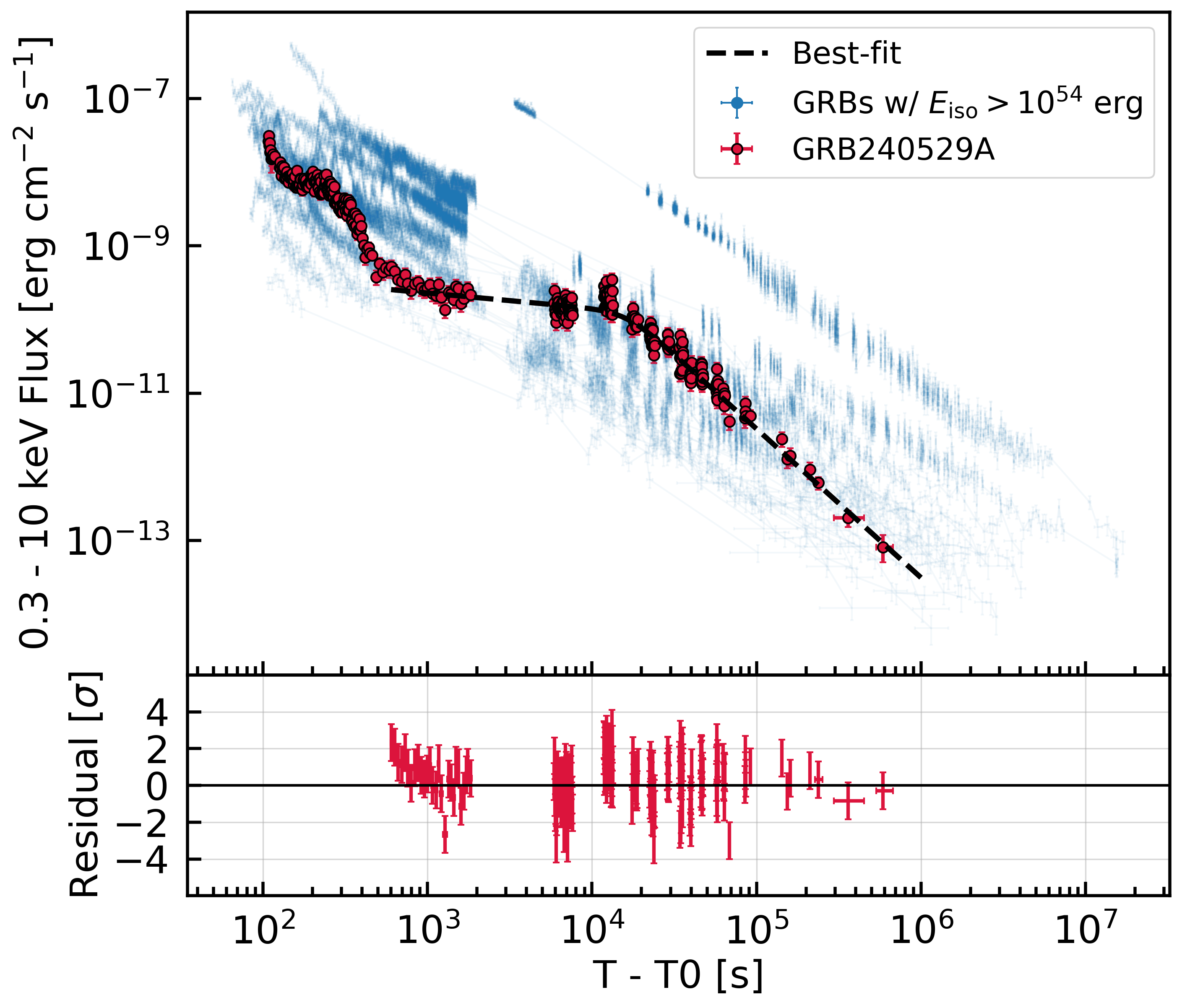}{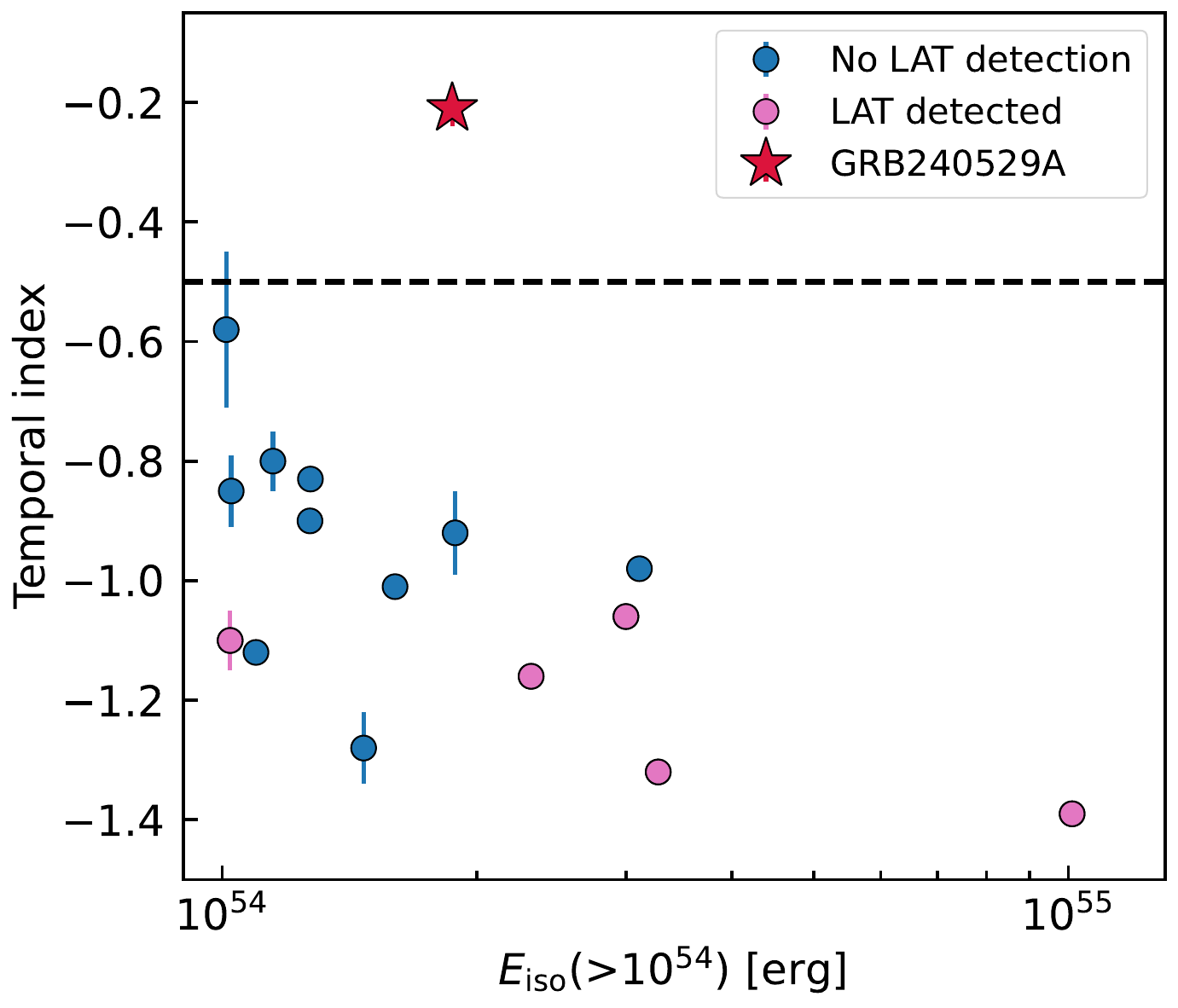}
    \caption{Left: 0.3--10~keV light curve of GRB~240529A (red) plotted with light curves of GRBs with $E_{\rm iso}>10^{54}$~erg (blue).
    The black dashed line is the smoothly broken power-law (SBPL) function fit to the XRT data after $t = \SI{600}{\second}$. 
    See Section \ref{subsec:xrt} for more details.
    Bottom panel shows the residuals of the difference in the best-fit model and data for GRB~240529A in units of sigma.
    Right: temporal indices of each burst plotted against $E_{\rm iso}$. The temporal indices are obtained by fitting with SBPL function, adopting the index before the break. GRB~240529A is marked as red star and GRBs detected by \textit{Fermi}-LAT \citep{Ajello_2019} are marked in pink. Typical value of the shallow decay index is marked by the black dashed line. $E_{\rm iso}$ values are calculated in the rest frame 1 keV--10 MeV range.
    These panels illustrate that the X-ray light curves of hyper-energetic GRBs often lack a shallow decay phase with a typical decay index as reported by \citet{Yamazaki_2020}.
    Samples are taken from \citet{Xue_2019, Sharma_2021, Gupta_2022, Rossi_2022, Mei_2022, Lesage_2023, Ror_2024}.}
    \label{fig:xray_lc}
\end{figure*}

The X-Ray Telescope (XRT; \cite{Burrows_2005}) on \textit{Swift} started observations of GRB~240529A \SI{107}{\second} after BAT trigger \citep{GCN36564}. 
The XRT light curve of GRB~240529A is presented in Figure~\ref{fig:xray_lc}.
From its light curve, the afterglow shows shallow decay expressed by a temporal index of $-0.21 \pm 0.03$, followed by a steeper decay than usual cases with a temporal index of $-2.02 \pm 0.04$ with a break at $\SI{1.63(0.05)e4}{\second}$ (see Section \ref{subsec:xrt} for the detail of the light curve study).
0.3--10 keV energy flux during the shallow decay phase is approximately \SI{2e-10}{erg\,\cm^{-2}\,\second^{-1}}, which is one of the brightest shallow decay in X-rays \citep{Ding_2022, Zhao_2019, Tang_2019}.
There is a bump in the X-ray lightcurve a few hundred seconds after the BAT trigger. 
While this can be explained by an X-ray flare, which is frequently seen in X-ray lightcurves \citep{Chincarini_2007}, \citet{Tian_2025} proposes a strange star model for this bump. 
We do not focus on this behavior in this paper.

An optical counterpart whose the position is consistent with the enhanced XRT position was detected by the \textit{Swift}/UVOT \citep{GCN36562}, which was later confirmed by other optical telescopes \citep{GCN36559, GCN36561, GCN36563, GCN36568, GCN36569, GCN36573, GCN36574, GCN36575, GCN36576, GCN36577, GCN36579, GCN36582, GCN36585, GCN36589, GCN36592, GCN36597, GCN36599, GCN36601, GCN36603, GCN36654, GCN36655, GCN36734, GCN37612}. 
It is noteworthy that an optical transient was clearly detected during the prompt phase by MASTER \citep{GCN36603}.
The GTC telescope reported a measurement of the redshift $z=2.695$ \citep{GCN36574}.

The radio detection of the afterglow was reported by couple of instruments approximately \SI{5}{days} after the burst, whose location is consistent with the optical afterglow position \citep{GCN36636, GCN36642}.

Given the fluence measurement of $S_{\gamma} = 1.3^{+0.4}_{-0.5}\times\SI{e-4}{erg\,\cm^{-2}}$ (20\,keV $-$ 10\,MeV) by Konus-\textit{Wind} and the redshift, the isotropic-equivalent gamma-ray energy of the prompt emission, $E_{\rm \gamma,iso}$ is estimated to be $E_{\rm \gamma,iso}=2.2^{+0.7}_{-0.8}\times\SI{e54}{erg}$ \citep{GCN36584}, which makes the GRB an energetic event among GRB population.

This means that GRB~240529A is a rare event with a clear X-ray shallow decay feature, with $E_{\rm \gamma,iso}$ exceeding \SI{e54}{erg}.
This is also illustrated by the right panel of Figure~\ref{fig:xray_lc}, which shows the distribution of the temporal index obtained from the light curve fit along with $E_{\rm iso}$.
GRB 240529A possess the most shallow decay index of 0.2 in the sample, while other bursts show steeper decay index than the typical value of 0.5, making GRB 240529A a distinct one among the sample.
%
%
Due to its unique characteristics, GRB~240529A may offer insights into the physical mechanism behind the shallow decay phase.


In this work, we present the \textit{Fermi}-LAT analysis of GRB~240529A, along with the X-ray analysis, which indicates a discovery of GeV emission associated with the shallow decay phase. 
The multi-wavelength modeling of the shallow decay emission is also presented with the aim of unraveling the physical mechanism underlying the shallow decay phase.

\section{Analysis and Results} 
\label{sec:analysis}

\subsection{\textit{Fermi}-LAT Analysis} 
\label{subsec:lat}

The \textit{Fermi}-LAT \citep{Atwood_2009} is sensitive to gamma rays with energies from 100 MeV to 300 GeV \citep{Abdollahi_2020} and has been continuously surveying the entire sky every 3 hr since beginning operation in 2008 August. 
We use \SI{6e5}{\second} of data of Pass 8 SOURCE class data \citep{Atwood_2013, Bruel_2018} between 100 MeV and 10 GeV. 
The upper bound of the energy is chosen to be 10 GeV, since an absorption by extragalactic background light (EBL) starts to be non-negligible in the higher energies.
Events within $20^{\circ}$ of the GRB position (enhanced XRT position: R.A., decl.\,= $335^{\circ}\!\!.\,358, +51^{\circ}\!\!.\,562$) are extracted.
Events detected at zenith angles larger than $100^{\circ}$ were excluded to limit the contamination from gamma rays generated by cosmic ray interactions in the upper layers of Earth’s atmosphere.

We performed an unbinned likelihood analysis with the latest Fermitools package (v2.2.0), utilizing the \verb|P8R3_SOURCE_V3| instrument response function. 
The region of interest (ROI) is modeled from the latest \textit{Fermi}-LAT source catalog based on 14 yr of data, 4FGL (data release 4, \cite{Ballet_2020}) for point and extended sources that are within $15^{\circ}$ of the ROI center, as well as the latest Galactic diffuse and isotropic diffuse templates (\verb|gll_iem_v07.fits| and \verb|iso_P8R3_SOURCE_V3_v1.txt|, respectively). 
Throughout the analysis, the spectral parameters of the 4FGL sources were kept fixed to the values from the catalog.

\begin{figure*}[t]
    \plottwo{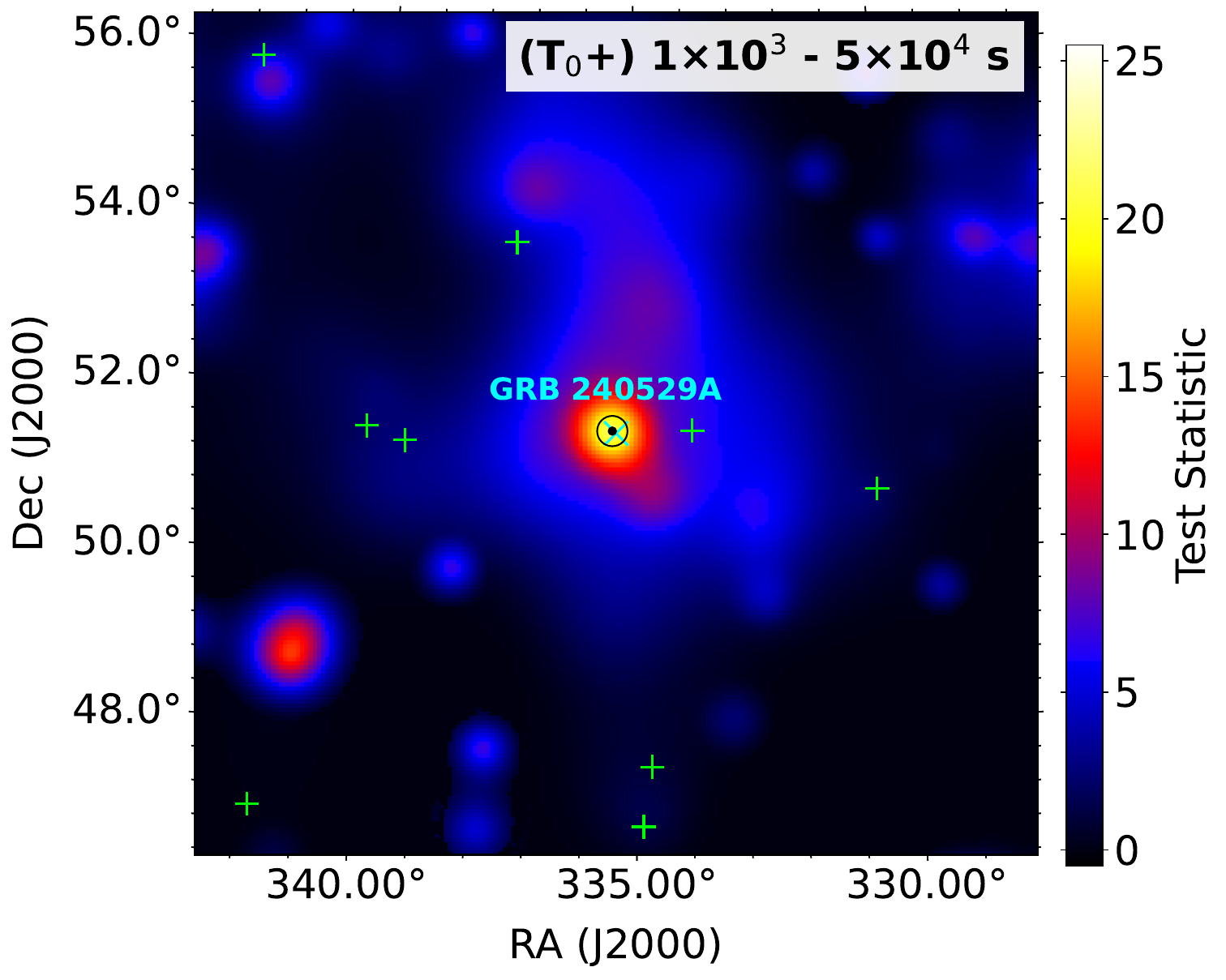}{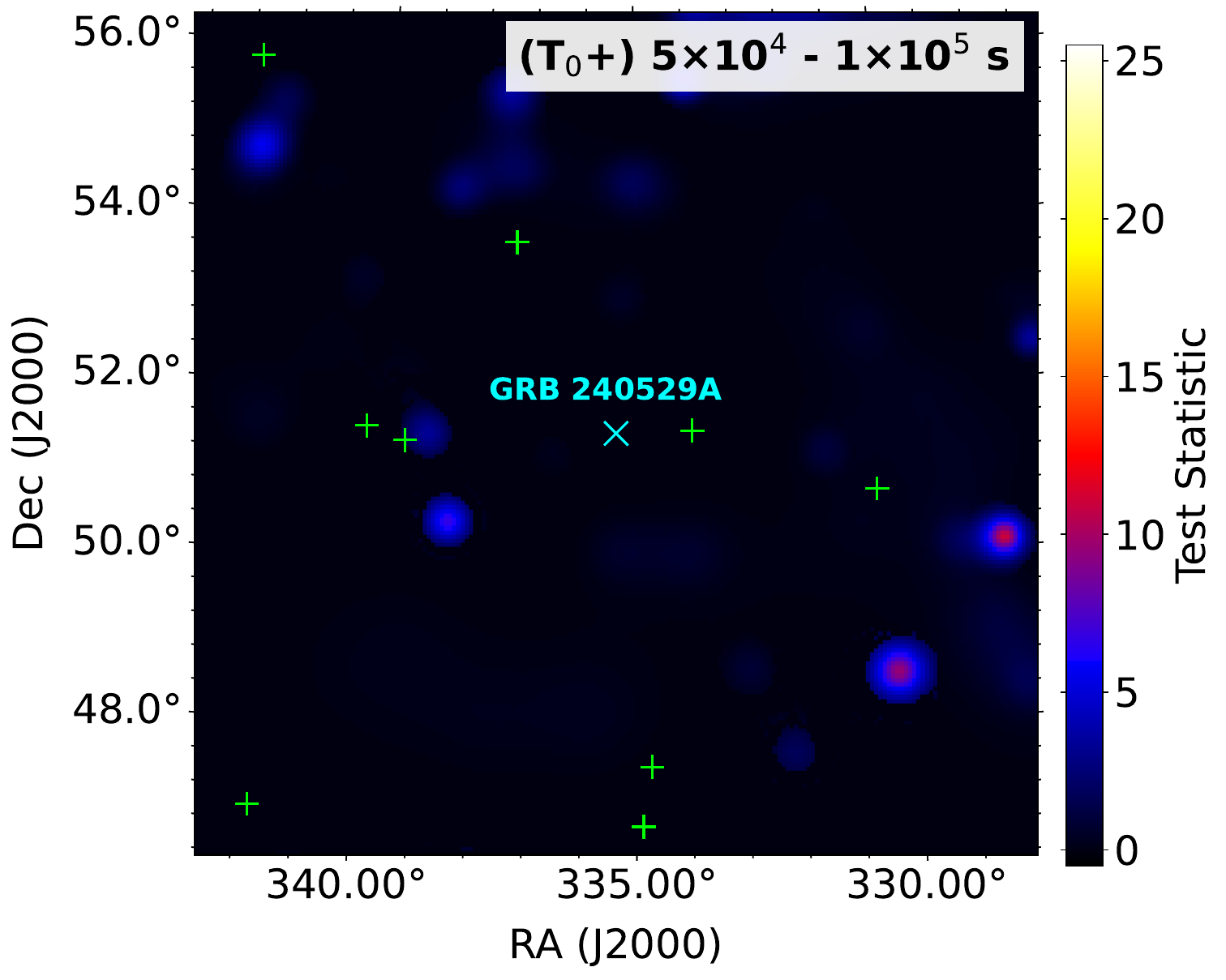}
    \caption{$10^\circ \times 10^\circ$ TS maps computed from \textit{Fermi}-LAT 100~MeV -- 10~GeV data with the enhanced XRT position (cyan cross) of GRB~240529A in the center. The pixel size is $0^{\circ}\!\!.\,05$ per pixel.
    Time intervals are \SI{1e3}{\second}\,$-$\,\SI{5e4}{\second} (left) and \SI{5e4}{\second}\,$-$\,\SI{1e5}{\second} (right).
    Locations of nearby 4FGL sources are also marked with green crosses in the figures.
    The LAT best-fit position and its error circle are marked in black for the left figure. 
    The error circles from BAT and optical instruments are consistent with the XRT position within the pixel size. 
    Note that there were no gamma-ray events between $T_0$ and $t = \SI{1e3}{\second}$ due to the ROI outside of LAT's field of view.}
    \label{fig:tsmap}
\end{figure*}

The background components were first determined using the data up to \SI{6e5}{\second} since $T_0$. 
The obtained best-fit spectral parameters of the background components were fixed and used for the further analysis.
Then, we computed count and test statistic (TS) maps in order to search for any residual gamma-ray emission. 
The TS value is defined to be the natural logarithm of the ratio of the likelihood of one hypothesis (e.g., presence of one additional source) and the likelihood for the null hypothesis (e.g., absence of source).
Note that there were no gamma-ray events between $T_0$ and $t = \SI{1e3}{\second}$ due to the ROI outside of LAT's field of view, and hence the maps were generated with time intervals after $t = \SI{1e3}{\second}$.
The computed TS maps are shown in Figure~\ref{fig:tsmap}.
The first time interval (\SI{1e3}{\second}--\SI{5e4}{\second}) is chosen to cover the X-ray shallow decay phase with a sufficient margin of time.
From the TS maps, we discovered a point-like gamma-ray source coincident in position with GRB~240529A in the time range of \SI{1e3}{\second}--\SI{5e4}{\second}.
This is the first GeV discovery for the shallow decay GRBs with $E_{\rm iso}>\SI{e54}{erg}$ with a typical shallow decay.
No significant source is found at times beyond $t = \SI{5e4}{\second}$.

\begin{figure}[]
    \epsscale{1.1}
    \plotone{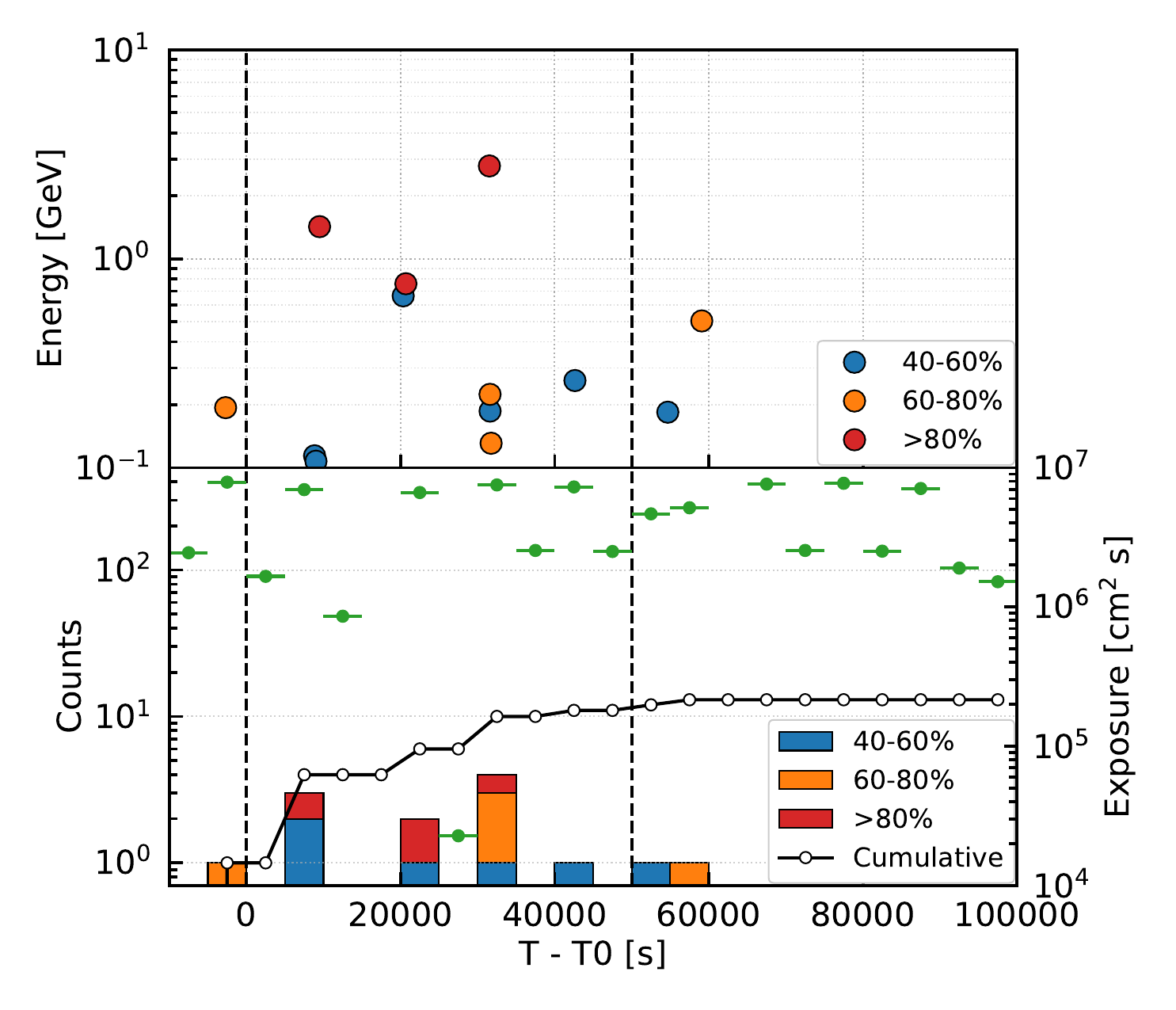}
    \caption{Top panel: The energy of each photon as a function of time. The events are categorized into three groups based on the probability of the GRB association: 40\%--60\% (blue), 60\%--80\% (orange), and $>$80\% (red).
    Bottom panel: The number of counts in each 5000\,s interval bin, with events grouped as shown in the top panel.
    The black line with white circle markers is the cumulative evolution of the counts with the probability greater than 40\%.
    The green markers show the exposure for each time bin.
    The vertical dashed lines in both panels represent the $T_0$ and $t = 5\times10^4$\,s, which the latter corresponds to the end time of the time interval used to compute the TS map shown in Figure~\ref{fig:tsmap} (left).
    Note that there is no event with probability between 80\% and 90\%, hence the probability greater than 80\% is effectively greater than 90\%.
    }
     \label{fig:lat_events}
\end{figure}

The probability of each photon being associated by GRB~240529A was investigated.
We first performed spectral likelihood analysis adopting \verb|PowerLaw2| model with spectral index fixed to 2.0, and then calculated the probability using \verb|gtsrcprob|.
The top panel of Figure~\ref{fig:lat_events} shows the energy of each photon divided into different probability ranges.
It can be seen that all events with probability greater than 80\% are included within $(1-3)\times10^4$\,s, overlapping the X-ray shallow decay phase.
Note that there is no event with probability between 80\% and 90\%, hence the probability greater than 80\% is effectively greater than 90\%.
The number of events after $T_0$ with probability greater than 90\% (40\%) is 3 (12), while no events are detected with probability greater than 40\% for off-source positions offset from the GRB position by $5^{\circ}$ (see Appendix~\ref{sec:appendix}).
Among the 3 events with probability greater than 90\%, the highest energy is 2.8 GeV, which was detected \SI{3.2e4}{\second} after $T_0$.
The bottom panel of Figure~\ref{fig:lat_events} shows the number of detected photons together with the exposure.
The number of photons increases during the first $3\times10^4$\,s, with no further increase observed afterward.
The cumulative evolution in the figure shows that the number of events with probability greater than 40\% reaches 10 around $3\times10^4$\,s after $T_0$.
The green markers represent the exposure time, which illustrates the sufficient exposure time for the time intervals with high-probability photons.
All the events with probability greater than 90\% are located within $0^{\circ}\!\!.\,2$ from the GRB position.

We localized the point source with \verb|gtfindsrc| to find the best-fit position and uncertainty.
The localized position for the new gamma-ray source is offset by $0^{\circ}\!\!.\,05$ from the exact position of GRB~240529A (R.A., decl.\,= $335^{\circ}\!\!.\,358, +51^{\circ}\!\!.\,562$) and has R.A., decl. = $335^{\circ}\!\!.\,43, +51^{\circ}\!\!.\,59$ (J2000). 
The corresponding 95\% positional uncertainty radius is $r = 0^{\circ}\!\!.\,18$, indicating a good consistency between the GRB position and the localized position.

To model the gamma-ray emission coincident with GRB~240529A, we added a point source at the GRB position to the source model. 
We set the spectrum to \verb|PowerLaw2| model, with its photon index fixed to 2.0 or allowed to vary. 
The derived photon flux (between 100 MeV and 10 GeV) is listed in Table~\ref{tab:lat_results}.
The returned TS value for the case of the fixed photon index is 20, corresponding to a detection significance of $4.5\sigma$.

We further investigated the gamma-ray emission by performing the time-resolved spectral analysis, utilizing the same model function (similarly as stated above, the photon index was either fixed to 2.0 or allowed to vary).
For t~$> \SI{5e4}{\second}$, where no significant signal is observed, 95\% confidence level flux upper limits are computed assuming a spectral index of 2.0 or 2.5. 
The results of the spectral analysis are presented in Table~\ref{tab:lat_results}.
Despite the obtained photon indices tend to show soft values (e.g., $2.99 \pm 0.64$ in \SI{1e3}{\second}\,$-$\,\SI{1.6e4}{\second}), the corresponding errors are large thus best-fit values are compatible with generic 2.0.

In order to check the stability of the results, analysis is also performed using Multi-Mission Maximum Likelihood \verb|3ML| \citep{3ML} and \verb|gtburst| softwares. \verb|3ML| is only used for the spectral analysis of the LAT data, whereas \verb|gtburst| is used for the entire LAT analysis, including the event selection up to the spectral analysis.
The obtained flux values are consistent within the $1\sigma$ statistical uncertainties.

Following the procedure described in \textit{Fermi}-LAT Second Gamma-Ray Burst Catalog paper \citep{Ajello_2019}, we calculated the duration of LAT emission and fluence of GRB~240529A, and compared with the latest LAT GRB catalog \footnote{\url{https://heasarc.gsfc.nasa.gov/w3browse/fermi/fermilgrb.html}}. 
The duration is \SI{2.2e4}{\second}, which is the third longest among the catalog.
The 100 MeV--100 GeV fluence is \SI{1.7e-5}{erg\,\cm^{-2}} (\SI{1e3}{\second}\,$-$\,\SI{5e4}{\second}; index free), which is in the top 10\% of the fluence values for the extended emission. Despite of the weak signal, the fluence resulted in a relatively high value due to the long duration of the signal.


\begin{deluxetable*}{p{3cm} >{\centering\arraybackslash}p{2cm} >{\centering\arraybackslash}p{2cm} >{\centering\arraybackslash}p{4cm} >{\centering\arraybackslash}p{4cm}}
    \tablecaption{Summary of the Fermi-LAT Spectral Analysis \label{tab:lat_results}}
    \tablehead{
        \colhead{Time since T0 (s)} & \colhead{TS~(index free)} & \colhead{Index} & 
        \colhead{\makecell{Integral Flux (index fixed) \tablenotemark{\scriptsize a} \\ $10^{-7}$\,ph\,cm$^{-2}$\,s$^{-1}$}} & 
        \colhead{\makecell{Integral Flux (index free) \\ $10^{-7}$\,ph\,cm$^{-2}$\,s$^{-1}$}}
    }
    \startdata
        \SI{1.0e3}{}\,$-$\,\SI{5.0e4}{} & 20~(21) & $2.35\pm0.38$ & $3.83 \pm 1.60$ & $5.23 \pm 2.31$ \\
        \SI{1.0e3}{}\,$-$\,\SI{1.6e4}{} & 10~(13) & $2.99\pm0.64$ & $7.89 \pm 4.34$ & $13.64 \pm 6.13$ \\
        \SI{1.6e4}{}\,$-$\,\SI{5.0e4}{} & 12~(12) & $2.01\pm0.49$ & $2.87 \pm 1.60$ & $2.90 \pm 2.04$ \\
        \SI{5.0e4}{}\,$-$\,\SI{1.0e5}{} & $<1$ & $-$ & $<$\,0.889~(1.31)\tablenotemark{\scriptsize b} & $-$ \\
    \enddata
    \vspace{5pt} 
    \footnotesize{ \textbf{Notes.} The flux is computed for each time range if TS $> 4$; otherwise, the 95\% confidence level upper limit is calculated. All the errors reported above represent 1$\sigma$ uncertainties (statistical only).}
    \vspace{-3pt}
    \tablenotetext{\scriptsize a}{\footnotesize The spectral index is fixed to 2.0 in the fit procedure.}
    \vspace{-6pt}
    \tablenotetext{\scriptsize b}{\footnotesize The result obtained for the spectral index of 2.5 is also presented.}
\end{deluxetable*}

\subsection{Swift-XRT Analysis} 
\label{subsec:xrt}

Launched in November 2004, the \textit{Swift} satellite is specifically designed for the study of GRBs \citep{Gehrels_2004}. 
\textit{Swift}-XRT is an X-ray imaging spectrometer with an energy range of 0.3--10 keV \citep{Burrows_2005}. 
The automated XRT light curves are extracted from the Burst Analyser website \footnote{\url{https://www.swift.ac.uk/burst_analyser/01231488/}} \citep{Evans_2010}.
In order to check the temporal behavior of GRB~240529A, we fit the 0.3--10 keV light curve with a smoothly broken power-law (SBPL) function \citep{Liang_2007, Yamazaki_2020}
\begin{equation}
    f_{\mathrm{SBPL}}(t) = f_0 \left[ \left(\frac{t}{t_{\mathrm{b}}}\right)^{\omega\alpha_1} + \left(\frac{t}{t_{\mathrm{b}}}\right)^{\omega\alpha_2} \right]^{-1/\omega},
\end{equation}
where $\alpha_1$ and $\alpha_2$ are decay indices before and after the break, respectively, and $t_{\rm b}$ is a break time.
The lower time range of the fit is fixed to \SI{600}{\second}, and also the sharpness parameter $\omega$ is fixed to 3, otherwise the fit does not converge correctly.
The value of 3 is chosen to simplify the comparison between the previous studies \citep{Liang_2007, Zhao_2019, Yamazaki_2020}. 
The best fitting temporal indices are $\alpha_1 = 0.21 \pm 0.03$ and $\alpha_2 = 2.02 \pm 0.04$, and the best fitting break time is $t_{\rm b} = \SI{1.63(0.05)e4}{\second}$.
The post-break decay index differs from the typical value of the normal decay phase, which is discussed later in Section~\ref{sec:modeling}.
The results for different $\omega$ values are shown in Appendix~\ref{sec:appendix2}.
As presented in Appendix~\ref{sec:appendix2}, the values of $\alpha_1$ remain consistent within statistical uncertainties for different $\omega$ values. 
As for $t_{\rm b}$ and $\alpha_2$, the result for $\omega = 1$ yields the longer break time $t_{\rm b} = \SI{1.92e4}{\second}$ and the steeper post-decay index $\alpha_2 = 2.31$ compared to other cases, although this does not affect the overall discussion.
The above fit procedure was also performed on 10 keV light curve, and similar values are acquired: $\alpha_1 = 0.21 \pm 0.03$, $\alpha_2 = 2.04 \pm 0.04$, and $t_b = \SI{1.62(0.06)e4}{\second}$.

In the later analysis, \SI{1.6e4}{\second} is chosen as the end time of the shallow decay.
Figure~\ref{fig:xray_lc} shows the XRT $0.3 - 10$\,keV light curve with the best-fit SBPL funtion.

As shown in Figure \ref{fig:xray_lc}, the X-ray light curve in the post-shallow decay phase shows no break that could be associated with a jet break.
Therefore, we derived a lower limit of the second break time by fitting the X-ray light curve with a smooth triple power-law (STPL) function
\citep{Liang_2008}
\begin{equation}
    f_{\mathrm{STPL}}(t) = \left[ f_{\mathrm{SBPL}}(t)^{-\omega_2} + f_1(t)^{-\omega_2} \right]^{-1/\omega_2},
\end{equation}
where $\omega_2$ is the sharpness parameter of the second break at $t_{\mathrm{b},2}$, and
\begin{equation}
    f_1(t) = f_{\mathrm{SBPL}}(t_{\mathrm{b},2}) \left( \frac{t}{t_{\mathrm{b},2}} \right)^{-\alpha_3}.
\end{equation}
The sharpness parameter $\omega$ and $\omega_2$ were fixed to 3 in the fit.
The parameters of the SBPL component are the same as those obtained from the best-fit results of the SBPL function alone, and only $\alpha_3$ and $t_{\mathrm{b},2}$ are set free.
Utilizing the STPL function, we scan a likelihood profile of the second break time $t_{\mathrm{b},2}$, and calculated the 90\% confidence level lower limit of $t_{\mathrm{b},2}$.
We find the lower limit of \SI{2.94e5}{\second} for the second break time $t_{\mathrm{b},2}$.

The XRT spectrum is obtained using the time-sliced spectra tool \footnote{\url{https://www.swift.ac.uk/xrt_spectra/addspec.php?targ=01231488&origin=slicedResult}} \citep{Evans_2009}. 
The time intervals are chosen to cover the shallow decay phase (\SI{1e3}{\second}\,$-$\,\SI{1.6e4}{\second}). 
The XRT data are fit using XSPEC v12.12.1 with a power-law function $dN/dE = N_0(E/1\, \mathrm{keV})^{-\Gamma}$ and an absorption component \verb|TBabs| \citep{Evans_2009}. 
The fit statistic is defined to be \verb|C-statistic|, which is suitable for XRT data. 
When the hydrogen column density is set free, its best-fit value gives low value compared to the Galactic value
$N_\mathrm{H}$ = \SI{4.06e21}{\per\cm\squared} \citep{Willingale_2013}, thus the column density is fixed to the Galactic value.
The final photon index obtained is $\Gamma = 2.13^{+0.04}_{-0.03}$ ($1\sigma$ error).

\subsection{Optical Data}
\label{subsec:opt}

Optical data are collected from public GCN circulars \citep{GCN36559, GCN36561, GCN36562, GCN36568, GCN36574, GCN36575, GCN36576, GCN36577, GCN36734}.
In the paper, we used data from R, r, L bands which have similar wavelength ranges.
Note that the L band used here is the one defined by GOTO \citep{Steeghs_2022}, which cover wavelength of $400 - 700$\,nm. 
Since the collected data were not corrected for the galactic extinction, we applied corrections following \citet{Cardelli_1989}.
Adopted values of band extinction are $A_\mathrm{R} = 0.73, A_\mathrm{r} = 0.78, A_\mathrm{L} = 0.88$, with the reddening of $E(B - V) = 0.29$ \citep{Schlafly_2011}.
The extracted optical light curve shows the initial plateau-like feature until the peak at $\sim \SI{e4}{\second}$ followed by the power-law decay, which is consistent with \citet{GCN37612} and \citet{Sun_2024}.

\section{Modeling and Discussion} 
\label{sec:modeling}

\begin{figure*}[ht!]
    \plotone{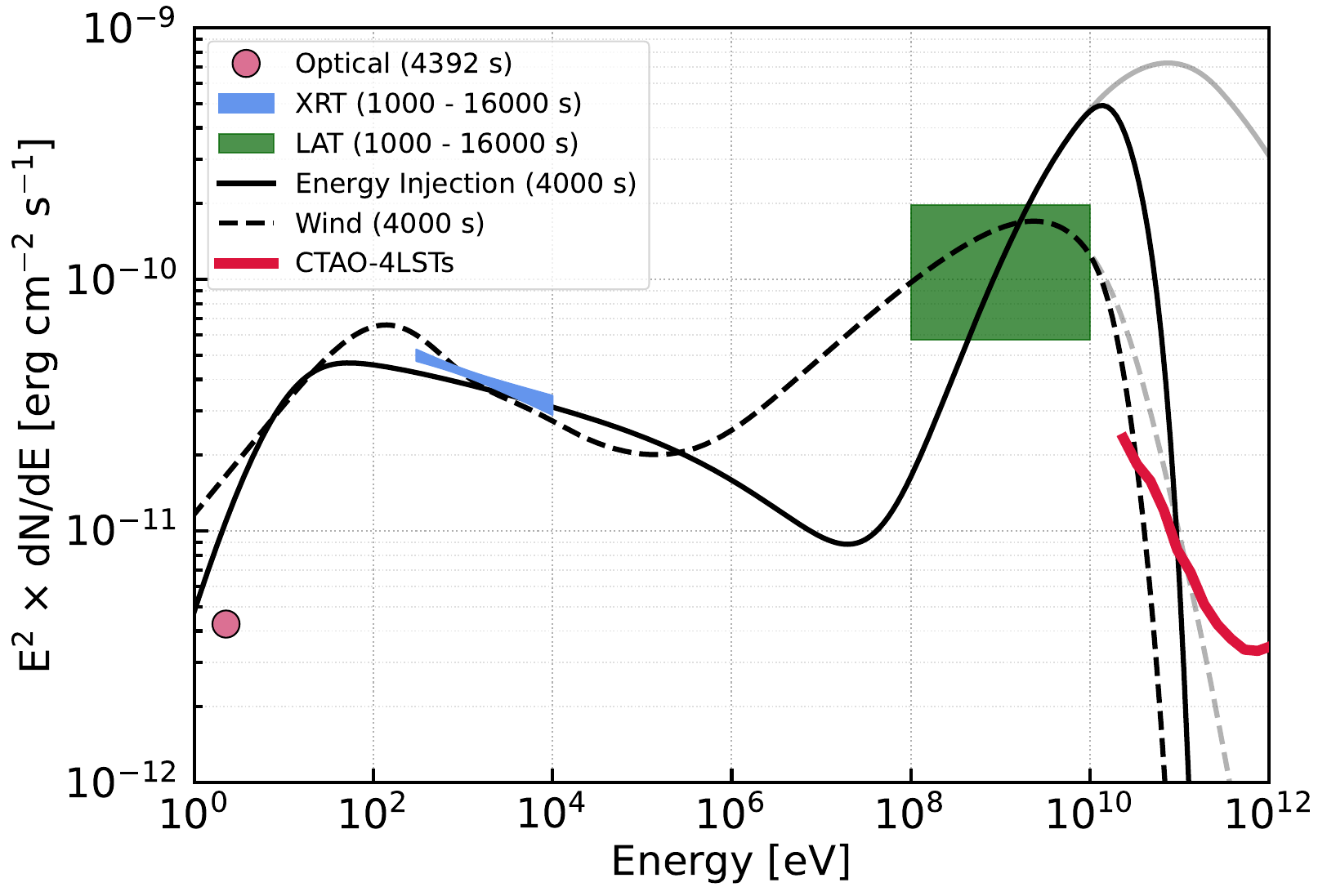}
    \caption{The multiwavelength SED at approximately 4000 s (see legend for the exact times) with modeling results of the energy injection model (solid line) and wind model (dashed line).
    The navy denotes XRT best-fit spectrum, the pink denotes the optical data, and the green denotes the LAT differential flux at 1 GeV calculated with the fixed spectral index of 2.0.
    Gray lines are model curves without EBL attenuation.
    The red line is the $5\sigma$ sensitivity curve of CTAO (only 4 LSTs) for observation time of $\SI{e4}{\second}$.}
     \label{fig:sed}
\end{figure*}

From \textit{Fermi}-LAT spectral analysis, the flux ratio between the GeV and keV energy ranges is $F_{\rm GeV}/F_{\rm keV} = 4.2 \pm 2.3$, strongly suggesting that the GeV emission arises from an inverse Compton component rather than a synchrotron component.
This is also supported by the photon index of 2.1 obtained from the XRT spectrum, which is difficult to explain the obtained LAT flux by the extrapolation of the XRT spectrum (see Figure~\ref{fig:sed}).
To produce an efficient inverse Compton emission, a high Compton parameter $Y$ is necessary, indicating a low energy fraction of the magnetic field $\epsilon_B$, or high energy fraction of electrons $\epsilon_{\rm e}$ and large blastwave energy $E_{\rm K,iso}$.

To reproduce the lightcurves and spectra, primarily focusing on the behavior during the shallow decay phase, we adopt the simulation code in \citet{2024ApJ...970..141A}, in which the shallow decay afterglow is discussed with time-dependent simulations. The code was developed in \citet{2017ApJ...844...92F} \citep[see also,][]{2020ApJ...905..105A}. The evolution of the electron and photon energy distributions in the shocked region is followed taking into account particle injection, synchrotron emission, synchrotron self-absorption, inverse Compton (IC) emission, $\gamma \gamma$-absorption, electron--positron pair creation, adiabatic cooling, and photon escape. The bulk Lorentz factor $\Gamma$ of the shocked region is self-consistently solved with the energy conservation law. The photon flux for observers is calculated considering the Doppler beaming and the surface curvature. As shown in \citet{2024ApJ...970..141A}, the photon spectra produced by this code are significantly different from the conventional analytical formulae \citep[see e.g.][]{2009ApJ...703..675N,2022MNRAS.512.2142Y}, in which the photon spectrum at an observer time $T$ is regulated by the local values of the circumstellar medium (CSM) density and $\Gamma$ with microscopic parameters (energy fraction of electrons $\epsilon_{\rm e}$, energy fraction of the magnetic field $\epsilon_B$, and the electron power-law index $p$) at a given radius $R \propto \Gamma^2 T$. Our results depend on the evolutionary history of the shock propagation.

Recent studies have shown that models with a single emission zone and constant microscopic parameters have difficulties reproducing multi-wavelength lightcurves including gamma-ray ones. Two-component models \citep[e,g.][]{2022MNRAS.513.1895R,2023MNRAS.522L..56S,2023ApJ...946L..23L,Sun_2024}, whose parameter number is twice that of the single-component model, have been proposed. Models with evolving microscopic parameters \citep[e.g.][]{2024ApJ...973L..44F} are also possible, and we cannot determine which model is likely.  Here, to discuss the mechanism of the shallow decay phase and avoid complicated model assumptions with many parameters, we focus on mainly the early part of the afterglow with the simple one-zone model. The later part of the afterglow, which may require another component, jet break, or the evolution of the model parameters, is out of our scope in this paper.

Here we adopt two models for the shallow decay phase: the energy injection model \citep{Zhang_2001, Zhang_2006, Granot_Kumar_2006} and the wind model \citep{Shen_2012,Dereli-Begue_2022}.
It should be mentioned that \citet{Sun_2024} proposed a model involving two shocks launched separately from the central engine in order to explain the multiwavelength emission of GRB~240529A
\footnote{From the results of \citet{Sun_2024}, the increase in energy in the second shock can be interpreted as a type of energy injection model, and the small initial Lorentz factor of the second shock is similar to the wind model.}.

In energy injection model, while the forward shock propagates in a CSM with a constant plasma density $0.07~\mbox{cm}^{-3}$, the isotropically-equivalent energy increases with $R$ as 
\begin{eqnarray}
E=E_{\rm i}+(E_0-E_{\rm i}) \left(\frac{R}{R_{\rm e}} \right)^2, \quad \mbox{for} \quad R<R_{\rm e},
\end{eqnarray}
where $E_{\rm i}=10^{53}$ erg, $E_0=2 \times 10^{55}$ erg, and $R_{\rm e}=2.5 \times 10^{18}$ cm.
The initial Lorentz factor is $\Gamma_0=300$. In this case, $\Gamma$ starts deceleration at $R\simeq 10^{18}$ cm with $\Gamma \propto R^{-0.6}$. 
For $R\geq R_{\rm e}$, the total energy is constant $E_0$ so that the standard Blandford-Mckee solution $\Gamma \propto R^{-3/2}$ is realized. The microscopic parameters are constant as $\epsilon_{\rm e}=0.2$, $\epsilon_B=8\times 10^{-5}$, and $p=2.8$.

In wind model, the CSM density decreases with $R$ as
\begin{eqnarray}
\rho=5.0 \times 10^{11} A_* \left( \frac{R}{1~\mbox{cm}} \right)^{-2}~\mbox{g}~\mbox{cm}^{-3},
\end{eqnarray}
where $A_*=2.0$,
while the total energy $E_0=10^{55}$ erg is constant. A small initial Lorentz factor $\Gamma_0=30$ is chosen to reproduce the shallow decay phase. 
The microscopic parameters are $\epsilon_{\rm e}=0.3$, $\epsilon_B=8\times 10^{-3}$, and $p=2.8$.


\begin{figure*}[ht!]
    \epsscale{1.1}
    \plottwo{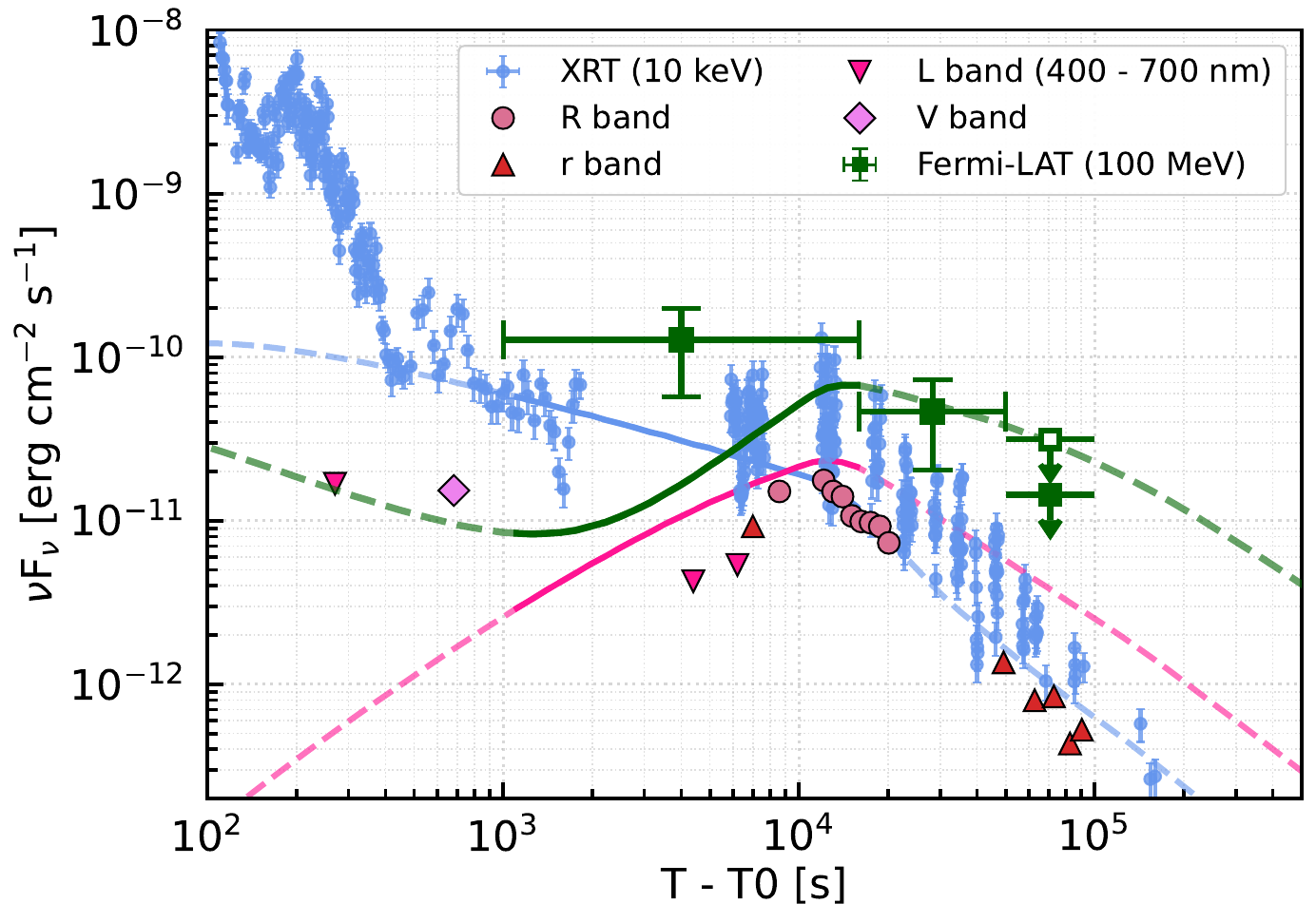}{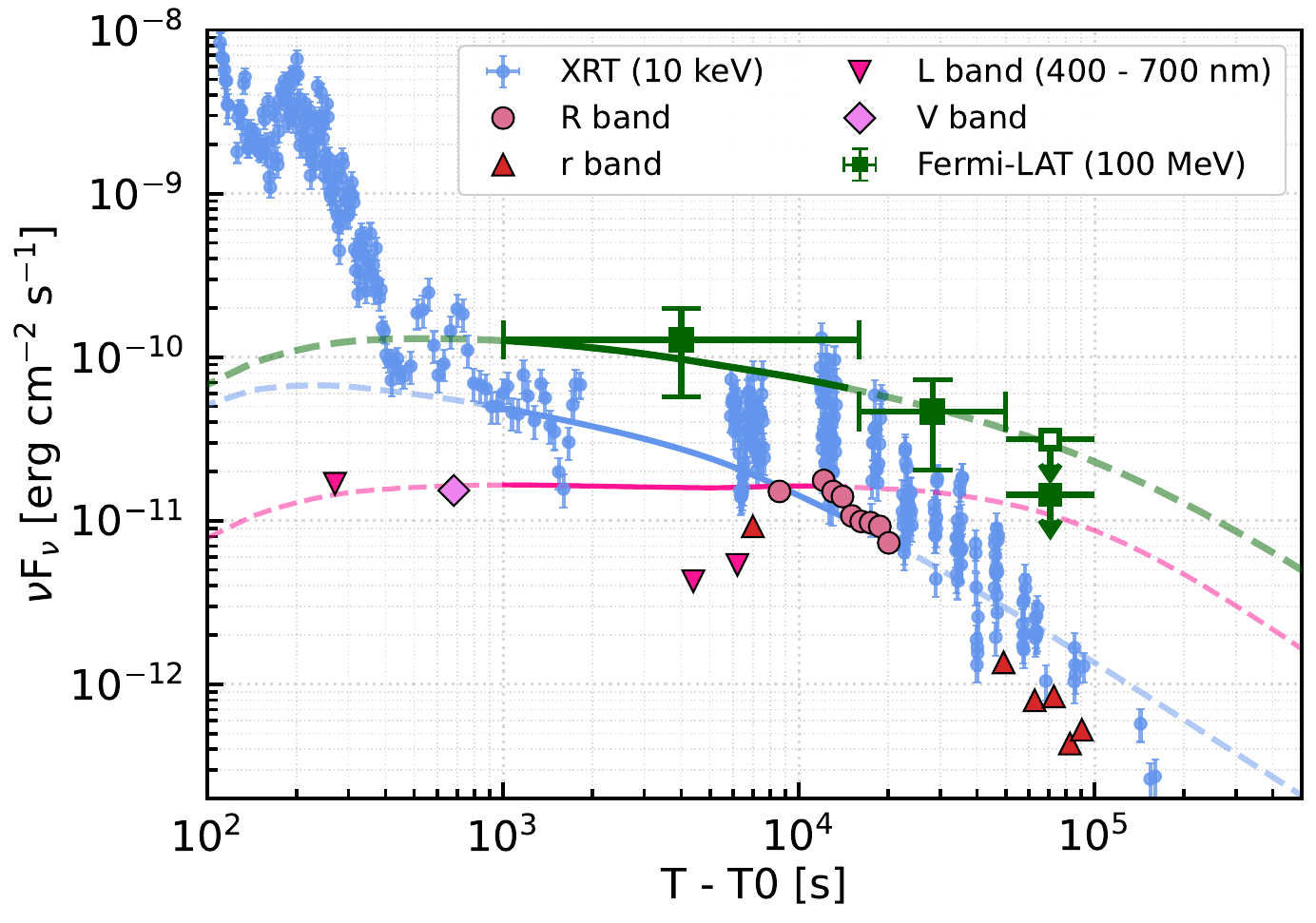}
    \caption{Multiwavelength light curves with modeling results for the energy injection model case (left) and the wind model case (right).
    The light blue points represent the XRT data, the green squares represent the LAT data, and the remaining colors denote optical data from different bands.
    The LAT 100 MeV light curve is converted from 100 MeV -- 10 GeV photon flux, assuming the photon index of 2.0.
    The LAT upper limits assuming the photon index of 2.5 are also shown as the green open squares.
    The 10 keV light curve of XRT is extracted from the Burst Analyser website.
    Note that the L band in the figures is the one defined by GOTO \citep{Steeghs_2022}.}
    \label{fig:mwl_lc}
\end{figure*}

Figure~\ref{fig:mwl_lc} and Figure~\ref{fig:sed} present reproduced multiwavelength light curves and spectrum, respectively, along with observational data.
Both models reproduce the X-ray flux and spectrum during the shallow decay phase ($T-T_0:600-1.6 \times 10^4$ s), with the X-ray flux consistent within a factor of 2\footnote{Although the models deviate from the X-ray data especially in the end of the shallow decay phase, the fine tuning of the model parameters would not greatly impact the main consequences (the interpretation of the GeV gamma-ray observations and the discussion on the jet energy) of this work.}.
Concerning the GeV light curve during the shallow decay phase, we compared the mean model flux, calculated between $(T_0+)\SI{1e3}-\SI{1.6e4}{\second}$, with the LAT data.
Both models yielded no significant differences with respect to the LAT data, showing less than 1\,$\sigma$ deviations.

In reproducing the GeV flux, the large $\epsilon_{\rm e}$ and small $\epsilon_B$ in the energy injection model are required to realize the observed GeV flux by synchrotron self-Compton (SSC) emission and the X-ray flux ($\propto \epsilon _{\rm e}^{p-1} \epsilon_B^{(p+1)/4}$) by synchrotron emission keeping the synchrotron peak frequency $\nu_{\rm m} \propto \epsilon_{\rm e}^2 \epsilon_B^{1/2}$ high enough to suppress the optical flux. In the energy injection model, the electron minimum Lorentz factor $\gamma_{\rm m}$ is much higher than that for the wind model. Therefore, the spectral peak frequency ($\sim 0.1$ TeV) of the SSC component is much higher than that of the wind model ($\sim1$ GeV). Therefore, the intrinsic TeV flux is suppressed in the wind model compared to the energy injection model.
Both models require $E_{\rm K,iso}$ exceeding \SI{e55}{erg}, which implies a significant amount of the blastwave kinetic energy, surpassing the radiation energy released during the prompt phase (\SI{2.2e54}{erg}).

As for GeV spectrum, the hard spectral index ($\sim1.24$) in the LAT energy range for the energy injection model is deviated from the LAT result ($2.99\pm0.64$) by $2.7\sigma$, while for the wind model case (index $\sim1.91$) the deviation is $1.7\sigma$.
Thus, the LAT data shows a tension with the energy injection model; however, the limited photon statistics of the LAT data prevent a definitive conclusion.
It should be noted that it is easier to reproduce the soft GeV spectrum in the wind model due to its low spectral peak frequency of the SSC component.

As demonstrated in \citet{2024ApJ...970..141A}, a precise measurement of the GeV light curve and the spectral evolution during the shallow decay phase is crucial to identify the physical mechanism behind the shallow decay.
Since the predicted TeV flux of the inverse Compton emission differs by more than an order of magnitude between the models, a precise measurement of TeV flux by upcoming Cherenkov Telescope Array Observatory (CTAO) will be able to constrain the shallow decay models.
It should be noted that for the energy injection model, CTAO can well detect TeV emission from similar events even under a strong EBL attenuation for a redshift $z=2.7$ as shown in Figure \ref{fig:sed}.
This redshift value is significantly greater than the highest redshift of $z=1.1$ ever detected by ground-based TeV instruments \citep{MAGIC_GRB201216C}.

After the shallow decay phase, both the two models deviate from the X-ray data for $T-T_0>10^5$ s.
As is well known, the optical lightcurve is the hardest one to reproduce. If the optical data should be treated as upper limits, both the two models are inconsistent after the shallow decay phase.
It is challenging to explain all the data, including optical observations, with a single jet model. 
Unlike \citet{Sun_2024}, our focus is primarily on the behavior during the first few $10^4$\,s. In fact, \citet{Sun_2024} note that the former shock has significant uncertainties and cannot be thoroughly discussed, so their work mainly concentrates on the later shock after few $10^4$\,s. This can be regarded as a complementary approach to our work.

As the X-ray lightcurve shows a steeper decay than usual, a potential explanation is the occurrence of a jet break just after the shallow decay phase. For the electron injection index $p=2.8$ inferred from our modeling, the temporal decay slope of 2.0 for the post-shallow decay phase is steeper than the usual decay index $3(p-1)/4=1.35$ and shallower than the slope after the jet break $p$.
Although the modeling of the afterglow after the shallow decay phase is out of scope in this paper, a complicated model, such as the jet break with the continuous energy injection or the parameter evolutions (evolving $p$, etc.), may be required to reconcile with the Fermi upper limit, optical data, and the steep decay of the X-ray lightcurves.
It should be noted that the 3 GeV photon was triggered at $t \sim \SI{3e4}{\second}$, which almost corresponds to the end of the X-ray shallow decay phase. A rapid evolution of the parameters may be allowed after this time.

Based on the X-ray light curve analysis and the multiwavelength modeling, we evaluate the jet opening angle and hence the collimation-corrected jet energy.
\citet{Sun_2024} calculated the isotropic X-ray luminosity during the shallow decay phase and attributed it to the dissipation of the magnetar wind, discussing the rotational period of a possible newborn magnetar. On the other hand, we newly calculated the collimation-corrected total jet energy and compared with the maximum rotational energy of the magnetar.

For a constant density CSM as in the energy injection model, a half-opening angle of the GRB jet can be expressed as \citep{Sari_1999}
\begin{multline}
    \theta_{\mathrm{j}} \simeq 0.05 \left( \frac{t_{\mathrm{j}}}{2.9\times 10^5\, \mbox{s}} \right)^{3/8} 
    \left( \frac{1+z}{3.695} \right)^{-3/8} \\
    \times \left( \frac{E_{\mathrm{K,iso}}}{\SI{2e55}{erg}} \right)^{-1/8} 
    \left( \frac{n}{\SI{0.1}{\cm^{-3}}} \right)^{1/8} \mbox{rad},
\end{multline}
where $t_{\mathrm{j}}$ is the time of jet break in seconds, $n$ is the CSM density, and $E_{\mathrm{K,iso}}$ is the isotropic kinetic energy of the blastwave.

Based on the values inferred from the XRT analysis and the modeling, $t_{\rm j}\geq\SI{2.9e5}{\second}$, $E_{\rm K,iso}=\SI{2e55}{erg}$, and $n=\SI{0.1}{\cm^{-3}}$, the jet half-opening angle is estimated to be $\theta_{\rm j} \gtrsim \SI{0.05}{rad}$.

Now one can estimate the beam-corrected total jet energy as
\begin{equation}
    E_{\mathrm{j}} = (1 - \cos \theta_{\mathrm{j}}) (E_{\gamma,\mathrm{iso}} + E_{\mathrm{K,iso}}) \gtrsim \SI{2.3e52}{erg}.
\end{equation}
For a wind medium, the half-opening angle $\theta_{\rm j}$ can be expressed as \citep{Chevalier_2000}
\begin{multline}
    \theta_{\mathrm{j}} \simeq 0.04 \left( \frac{t_{\mathrm{j}}}{2.9\times 10^5\, \mbox{s}} \right)^{1/4} 
    \left( \frac{1+z}{3.695} \right)^{-1/4} \\
    \times  \left( \frac{E_{\mathrm{K,iso}}}{\SI{1e55}{erg}} \right)^{-1/4} 
    \left( \frac{A_*}{\SI{2.0}{\cm^{-3}}} \right)^{1/4} \mbox{rad},
\end{multline}
hence the beam-corrected jet energy $E_{\rm j} \gtrsim \SI{1.2e52}{erg}$.
Here $t_{\rm j}\geq\SI{2.9e5}{\second}$, $E_{\rm K,iso}=\SI{1e55}{erg}$, $A_*=\SI{2.0}{\cm^{-3}}$, and $\theta_{\rm j}\gtrsim\SI{0.04}{rad}$ are adopted.
Note that given the weak dependence of medium density on the jet opening angle for both cases, the derived value of the jet opening angle is not sensitive to the assumed medium density.

For both cases the beam-corrected jet energy $E_{\rm j}$ is close to the maximum rotational energy of the magnetar (1--2$\times 10^{52}$ erg; \cite{Sharma_2021, Li_2018}).
If some fractions of the rotational energy is converted to the jet energy, then a magnetar may be difficult to power the jet.
In such case, a black hole may be favored as the central engine.
However, the magnetar is still possible as the central engine, which in this case implies a high efficiency of the energy conversion from the rotational energy into the jet energy.
The estimated jet energy will be reduced if the earlier jet break at $t \sim \SI{3e4}{\second}$ is considered as discussed above. Adopting the same parameters for the blastwave's kinetic energy $E_{\rm K,iso}$ and ambient medium density $n(A_*)$, the beam-corrected jet energy $E_{\rm j}$ will reduce to $\SI{4.1e51}{erg}$ and $\SI{3.7e51}{erg}$ for the energy injection model and the wind model, respectively.

It should be emphasized that GRB~240529A is a unique burst, which shows the energetic prompt phase followed by the clear X-ray shallow decay phase.
The rarity of the X-ray shallow decay in hyper-energetic GRBs (e.g., \citet{Sharma_2021, Yamazaki_2020, Ding_2022}) may indicate different properties of the central engine or surrounding environments compared to those of less energetic GRBs.
Since other hyper-energetic GRBs do not need the evident shallow decay phase, this may imply that, for majority of hyper-energetic GRBs, the energy injection into the shockwave from the central engine may be instantaneous, or the initial Lorentz factor of the launched jet is high.

\section{Summary and Conclusion} 
\label{sec:summary}

X-ray light curves of gamma-ray bursts afterglow exhibit a variety of features, with the shallow decay phase being the most enigmatic.
Various models have been proposed to explain the shallow decay phase, but the physical mechanism of the shallow decay remains a mystery.
Previous studies argue that hyper-energetic or GeV--TeV detected GRBs display no or less prominent shallow decay phase in their X-ray light curves \citep{Yamazaki_2020, Sharma_2021, Ding_2022}.

Recently discovered GRB~240529A has an extremely large isotropic gamma-ray energy release of \SI{2.2e54}{erg} and a prominent X-ray shallow decay phase with a decay index of 0.2.
This indicates that the burst is a unique case compared to previous studies and may provide a valuable key to understanding the physical process of the shallow decay phase.
GeV--TeV observations will be crucial for distinguishing between different models of the shallow decay \citep{2024ApJ...970..141A}.


In this work, we presented \textit{Fermi}-LAT analysis of GRB~240529A, indicating the evidence of GeV emission ($4.5\sigma$) covering the X-ray shallow decay phase of the GRB.
This can be interpreted as the first case for hyper-energetic GRBs with a typical shallow decay phase.
The flux level of the GeV emission is $4.2\pm2.3$ times larger than the keV flux. 
Combined with the XRT spectral index of 2.1, this suggests that the GeV emission is dominated by strong IC emission.
In order to discuss in a multiwavelength context, we performed the multiwavelength modeling based on the time-dependent simulations and tested the two promising models of the shallow decay phase: the energy injection model and the wind model.
In the modeling, small energy fraction of the magnetic field, and large values of electron energy fraction and blastwave energy are adopted for both models to account for the efficient IC emission.
As a result, the blastwave energy exceeding \SI{e55}{erg} is required, which implies a significant amount of the blastwave kinetic energy, surpassing the radiation energy released during the intense prompt phase.
Together with XRT analysis done in this work, the modeling results suggest that the lower bound of the collimation-corrected total jet energy is (1--2)$\times10^{52}$~erg.
This value is close to the maximum possible rotational energy of the magnetar, indicating either a black hole or a magnetar with extremely high jet energy conversion efficiency as the central engine of GRB~240529A.
Both models reproduce the X-ray and GeV data during the shallow decay phase, though the GeV spectrum indicates some tension with the energy injection model.
Although distinguishing between them requires better statistics, our modeling shows that gamma-ray observations are a useful tool in constraining the shallow decay models.
Future GeV$-$TeV observations by CTAO will play a pivotal role in constraining models of the shallow decay phase.

We appreciate the anonymous referee for the helpful advice.
This research has made use of Fermi data obtained through High Energy Astrophysics Science Archive Research Center Online Service, provided by the NASA/Goddard Space Flight Center. 
This work made use of data supplied by the UK Swift Science Data Centre at the University of Leicester.
This research has made use of the CTA instrument response functions provided by the CTA Consortium and Observatory, see https://www.ctao-observatory.org/science/cta-performance/ (version prod5 v0.1; \cite{CTAO}) for more details.
This work is supported by a Grant-in-Aid for JSPS Fellows Grant Nos. 23KJ1222 (KT) and 23KJ2094 (TO), and KAKENHI grant Nos. 22K03684, 23H04899, and 24H00025 (KA).

%

\vspace{5mm}


\software{
    Fermitools: Fermi Science Tools \citep{Fermitools},
    XSPEC \citep{XSPEC},
    matplotlib \citep{matplotlib},
    numpy \citep{numpy},
    APLpy \citep{aplpy2012, aplpy2019},
    Astropy \citep{astropy:2013, astropy:2018, astropy:2022},
    gammapy \citep{gammapy:2023, gammapy:zenodo-1.1}
    }



\appendix
 \restartappendixnumbering
 
\section{The GRB association probability for off-source events}
\label{sec:appendix}

\begin{figure}[]
    \epsscale{0.7}
    \plotone{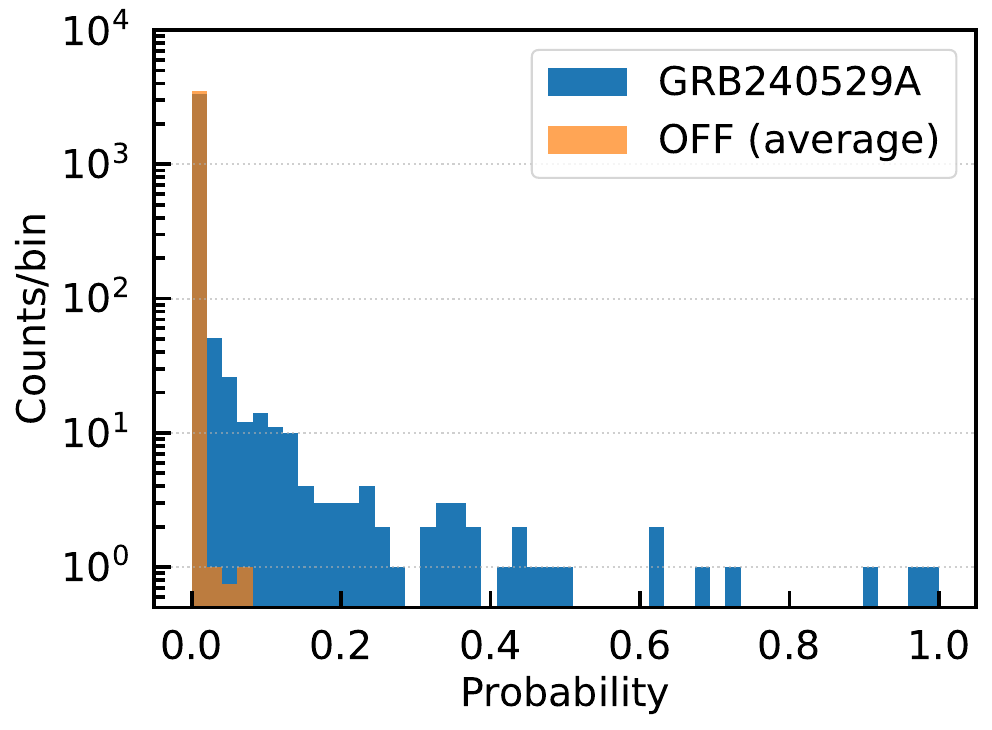}
    \caption{Distributions of the association probability for the GRB case (blue) and the off-source case (orange).
    The average of the 4 off-source positions are presented in the figure.}
     \label{fig:lat_prob_dist}
\end{figure}

In order to check the off-source dependence of the LAT events with probability higher than 40\%, we evaluated the probability distribution of the events being associated by off-source positions. The off positions are chosen to be the positions offset by $5^{\circ}$ from the GRB position. The $\pm 5^{\circ}$ offset is applied to the right ascension and the declination of the GRB coordinates, resulting in 4 off-source positions in total.
The association probability of the off positions is computed similarly as in the GRB case using \verb|gtsrcprob|, adopting \verb|PowerLaw2| model with the spectral index fixed to 2.0.
The obtained distribution of the association probability is presented in Figure~\ref{fig:lat_prob_dist}.
It shows that there are no events with the probability higher than 10\% for the averaged off distribution.
This indicates that the LAT analysis presented in this work is robust, with no contribution from background events.

\section{Details of temporal analysis of XRT data}
\label{sec:appendix2}

Table~\ref{tab:xrt_results} displays the results of XRT temporal analysis for different sharpness parameter $\omega$.
\citet{Tian_2025} fitted the X-ray light curve of GRB~240529A with superposition of one power-law and two smoothly broken power-law models, and obtained slightly shallower decay index before the break ($0.09\pm0.04$) compared to our results. The other temporal parameters are relatively unaffected and consistent with ours. It should be mentioned that they fixed the sharpness parameter to $\omega = 7$ (for the plateau phase on which we are focusing), thus our results obtained with $\omega = 5$ are the most compatible compared to those with $\omega$~=~1, 2, 3, and 4.

\begin{deluxetable*}{
    >{\centering\arraybackslash}p{2cm} 
    >{\centering\arraybackslash}p{3cm} 
    >{\centering\arraybackslash}p{3cm} 
    >{\centering\arraybackslash}p{3cm}
    }
    \tablecaption{Results of XRT Temporal Analysis for Different Sharpness Parameter $\omega$ \label{tab:xrt_results}}
    \tablehead{
        \colhead{$\omega$} &
        \colhead{\makecell{$t_{\rm b}$ \\ (\SI{e4}{\second})}
        }& 
        \colhead{$\alpha_1$} &
        \colhead{$\alpha_2$}
    }
    \startdata
        1 & $1.92 \pm 0.10$ & $0.15 \pm 0.03$ & $2.31 \pm 0.06$ \\
        2 & $1.70 \pm 0.06$ & $0.20 \pm 0.03$ & $2.09 \pm 0.04$ \\
        3 & $1.63 \pm 0.05$ & $0.21 \pm 0.03$ & $2.02 \pm 0.04$ \\
        4 & $1.59 \pm 0.05$ & $0.21 \pm 0.03$ & $1.99 \pm 0.03$ \\
        5 & $1.57 \pm 0.05$ & $0.21 \pm 0.03$ & $1.97 \pm 0.03$ \\
    \enddata
    \footnotesize{ \textbf{Notes.} The parameter $\omega$ represents the sharpness factor. $t_{\rm b}$ is the break time in units of $10^4$ seconds. $\alpha_1$ and $\alpha_2$ denote the temporal decay indices before and after the break, respectively. All the errors reported above represent 1$\sigma$ uncertainties (statistical only).}
\end{deluxetable*}


\bibliography{reference}{}
\bibliographystyle{aasjournal}



\end{document}